\documentclass[sn-mathphys,pdflatex]{sn-jnl}
\begin{document}

\title[Searches for QCD instantons with forward proton tagging]{Searches for QCD instantons with forward proton tagging}

\author*[1]{\fnm{Marek} \sur{Tasevsky}}\email{Marek.Tasevsky@cern.ch}
\author[2]{\fnm{Valery} \sur{Khoze}}\email{v.a.khoze@durham.ac.uk}
\author[2]{\fnm{Daniel} \sur{Milne}}\email{daniel.l.milne@durham.ac.uk}
\author[3]{\fnm{Michail} \sur{Ryskin}}\email{ryskin@thd.pnpi.spb.ru}

\affil*[1]{\orgname{Institute of Physics of the Czech Academy of Sciences}, \orgaddress{\street{Na Slovance 1999/2}, \city{Prague}, \postcode{18221}, \country{Czech Republic}}}

\affil[2]{\orgdiv{Department of Physics}, \orgname{University of Durham}, \city{Durham}, \postcode{DH1 3LE}, \country{United Kingdom}}

\affil[3]{\orgname{Petersburg Nuclear Physics Institute}, \orgaddress{\street{NRC ``Kurchatov Institute'', Gatchina}}, \city{St. Petersburg}, \postcode{188300}, \country{Russia}}    
 
\abstract{We study the possibility to observe heavy ($M_{\rm inst}> 60$~GeV) QCD instantons at the LHC in events with one
  or two tagged leading protons including fast simulation of detector and pile-up effects. We show that the expected
  instanton signal in a single-tagged configuration is strongly affected by central detector and pile-up
    effects. For double-tagged
  approach, where larger integrated luminosities and hence larger pile-up contaminations need to be considered, the
  combinatorial background overwhelms the expected signal. We suggest that additional time information about tracks at
  central and forward rapidities would be crucial for potential improvements.} 


\maketitle

\section{Introduction}

Instantons are non-perturbative classical solutions of Euclidean equations of motion in non-abelian gauge
theories~\cite{Belavin:1975fg}. In the semi-classical limit, instantons
describe quantum tunneling between different vacuum sectors of the theory \cite{tHooft:1976snw,Callan:1976je,Jackiw:1976pf}. They are either directly responsible for generating or at least contributed to many key aspects 
of non-perturbative low-energy dynamics of strong interactions~\cite{tHooft:1986ooh,Callan:1977gz,Novikov:1981xi,Shuryak:1982dp,Diakonov:1984vw,Schafer:1996wv}. 
These include the role of instantons in the breaking of the $U(1)_A$ symmetry and the spontaneous 
breakdown of the chiral symmetry, the formation of quark and gluon condensates, $<0\lvert\bar qq \rvert 0>$ and
$<0\lvert G^a_{\mu\nu}G^a_{\mu\nu}\rvert 0>$ and so on.

Unfortunately, up to now, the QCD instanton has never been observed experimentally.~\footnote{The QCD instanton
production in inclusive events at the LHC was considered in Refs.~\cite{Khoze:2019jta,Khoze:2020tpp,Amoroso:2020zrz}.}
The problem is that the large-size instanton is very challenging to distinguish from 
various possible soft QCD contributions, while the cross-section of the heavy (small-size) instanton production is 
exponentially suppressed by the $e^{-2S_I}$ factor, where the corresponding to instanton action $S_I=2\pi/\alpha_s$.
Nevertheless, it would be very appealing to observe the small-size instanton signal since in such a case, the
uncertainties due to the soft QCD and other non-perturbative effects are better controllable.\\

The characteristic signature of small-size instanton is the production of a large number of isotropically distributed
(mini)jets. That is, we are searching for the high multiplicity events with large (close to 1) sphericity $S$.\\
 
It was shown in Ref.~\cite{Sas:2021pup} that at high multiplicities the role of the multiple parton interactions (MPI)
strongly increases and the resulting sphericity of these MPI events becomes also close to one. Therefore, first of all,
we have to suppress the MPI contributions. This can be done by selecting the events with large rapidity gaps (LRGs). 
Indeed, it was demonstrated in Ref.~\cite{Khoze:2021jkd} that in this case there is a good chance to observe the
instanton signal at the LHC. Recall that the kinematics considered in Ref.~\cite{Khoze:2021jkd} corresponds to the
LRG events, selected by observing the leading proton which carries away a very close to 1 ($x_L\sim 0.997$) fraction of
its initial momentum in the ALFA or TOTEM detectors~\cite{AbdelKhalek:2016tiv,TOTEM:2008lue}.
Since the remaining energy is quite small, the mass of the instanton is not too large: $M_{\rm inst}\sim 20-40$~GeV. 

In the present paper we discuss the possibility to observe the higher mass instantons ($M_{\rm inst}>60$ GeV) by tagging
the leading protons with the dedicated forward proton detectors (FPDs): AFP~\cite{AFP,Tasevsky:2015xya} on the ATLAS
side or CT-PPS~\cite{CT-PPS,CMS:2021ncv} on the CMS side, when the remaining fraction of beam energy,
$\xi=1-x_L\sim 0.03$. They were both installed in Run~2 and first experience shows that they cover a
$0.02 < \xi < 0.15$ region. They are also equipped by time-of-flight (ToF) detectors with a time resolution of 10~ps
expected to be achieved in Run~4. The goal for the Run~3 data taking is 20~ps. 
  
Due to the strong, $e^{-2\pi/\alpha_s}$, suppression of a heavy instanton amplitude the expected cross-section becomes
rather small. Thus, we have to consider the possibility to work at a large luminosity and account for the pile-up
background. We study the 'one LRG' kinematics, where only one leading proton is detected, and the central instanton
production, when both  leading protons are observed.\\
  
We will follow here and below the results of the previous~\cite{Khoze:2021jkd,Khoze:2021pwd} papers.
In section~\ref{sec:QCDinst} we recall the definition and basic formulae for the instanton-induced processes.
In section~\ref{sec:generation} the main details of the generation and selection of instanton events with one or two
LRGs are discussed. Searching strategy and the optimal cuts are described ed in section~\ref{sec:strategy}, while
the results are presented in section~\ref{sec:results}. We conclude in section~\ref{sec:conclusions}.

\section{The QCD instanton}\label{sec:QCDinst}
In QCD, the instanton configuration consists of the gauge field,
\begin{eqnarray}
A_\mu^{a\, {\rm inst}}(x) &=& \frac{2 \rho^2}{g} \frac{\bar{\eta}^a_{\mu\nu} (x-x_0)_\nu}{(x-x_0)^2((x-x_0)^2+\rho^2)}\,,
\label{eq:instFT1}
\end{eqnarray} 
along with the fermion components for light ($m_f <1/\rho$) fermions,
\begin{equation}
\bar{q}_{Lf} = \psi^{(0)}(x) \,, \quad q_{Rf} = \psi^{(0)}(x) \,.
\label{EQ_i_gauge}
\end{equation}
The gauge field $A_\mu^{a\, {\rm inst}}$ is the Belavin-Polyakov-Schwartz-Tyupkin (BPST) instanton solution~\cite{Belavin:1975fg}  of the self-duality equations in the singular gauge. Here $\rho$ is the instanton size and $x_0$ is the instanton position. 
Constant group-theoretic coefficients $\bar{\eta}^a_{\mu\nu}$ are the 't Hooft eta symbols defined in Ref.~\cite{tHooft:1976snw}. The fermionic components $\psi^{(0)}$ are the corresponding normalised solutions of the Dirac equation
 $\gamma^\mu D_\mu[A_\mu^{a\, {\rm inst}}] \psi^{(0)} \,=0$. These are the fermion zero modes of the instanton.
The instanton configuration is a local minimum of the Euclidean action, and the action on the instanton is given by
$S_I= \frac{8\pi^2}{g^2} = \frac{2\pi}{\alpha_s}$.

The instanton configuration in Eq.~\eqref{eq:instFT1} has a topological charge equal to one and thus, due to the chiral anomaly, the instanton processes violate chirality. If the instanton is produced by a two-gluon initial state, the final state of this instanton-mediated process will have $N_f$ pairs of quarks and anti-quarks with the same chirality,
\begin{equation}
\label{e2}
g+g\to n_g\times g + \sum^{N_f}_{f=1}(q_{Rf}+\bar q_{Lf})\ ,
\end{equation} 
where $N_f$ is the number of light flavours relative to the inverse instanton size, $m_f <1/\rho$.
The instanton contribution to the amplitude for this process comes from expanding the corresponding path integral in the instanton field background.
At leading order in the instanton perturbation theory, the amplitude
takes the form of an integral over the instanton collective coordinates (see e.g. Refs.~\cite{Khoze:2019jta,Khoze:2020tpp} for more detail)
\begin{equation} 
{\cal A}^{\rm \, L.O.}_{\, 2\to\, n_g+ 2N_f} = \int d^4 x_0 \int_0^\infty d\rho \, D(\rho) \, e^{-S_I}\,
\prod_{i=1}^{n_g+2} A_{{\rm LSZ}}^{\rm inst}(p_i;\rho)\, \prod_{j=1}^{2N_f} \psi^{(0)}_{{\rm LSZ}}(p_j; \rho).
\label{EQ_ampgg}
\end{equation}
The factors of $A_{{\rm LSZ}}^{\rm inst}(p_i;\rho)$ and 
$ \psi^{(0)}_{{\rm LSZ}}(p_j; \rho)$ are the standard insertions of the LSZ-reduced instanton fields in the momentum representation and $D(\rho)$ is given by the known expression for the instanton density~\cite{tHooft:1976snw}.
\medskip

From the point of view of
Feynman graphs the leading order instanton amplitude (Eq.~\eqref{EQ_ampgg}) reveals itself as a family of multi-particle vertices (with different numbers of emitted gluons), integrated over the instanton position and size. It describes the
emission of a large number of gluons, $n_g \propto E^2/\alpha_s$, together with a fixed number of quarks and anti-quarks, one pair for each light flavour in accordance with Eq.~\eqref{e2}.
The semi-classical suppression factor, $\exp(-S_I)=\exp(-2\pi/\alpha_s)$, will be partially compensated by the growth with jet energy, $E$, of the high multiplicity cross-section for the process in Eq.~\eqref{e2}.
The fully factorised structure of the field insertions on the right-hand side of Eq.~\eqref{EQ_ampgg} implies that at leading order in instanton perturbation theory there are no correlations between the momenta of the external legs in the instanton amplitude. The momenta of individual particles in the final state are mutually independent, apart from overall momentum conservation.

Thus, to discover the QCD instanton we have to observe in the final state a multi-particle cluster or a fireball which contains in general a large number of isotropically distributed gluon (mini)jets accompanied by $N_f$ pairs of light quark jets generated by a subprocess such as in Eq.~\eqref{e2}. 

It is quite challenging, however, to identify the instanton on top of the underlying event. Recall, that the instanton is not a particle and there will be no peak in the invariant mass, $M_{\rm inst}$, distribution~\footnote{What we mean by the instanton mass is the partonic energy $\sqrt{\hat{s}}$ of the initial 2-gluon state in the process
  in Eq.~\eqref{e2}. As we integrate over the Bjorken x variables when computing hadronic cross-section, we sum over a broad range of instanton masses. }.
 The mean value of $M_{\rm inst}$ can at least in principle be ``measured" or reconstructed as the mass of the minijet system created by the instanton fireball in each given event. Talking about the instanton we actually mean a family of objects of different sizes, $\rho$ and different orientations in the colour and the Lorentz spaces. The mean value of $M_{\rm inst}$ depends on $\rho$, increasing when $\rho$ decreases. Since experimentally it is impossible to measure the instanton size, $\rho$, below we use the  mass $M_{\rm inst}$ to characterise the properties of the instanton production event.

\section{Generation of events}\label{sec:generation}
As discussed above, we expect a large `underlying event' background. However the background caused by multi-parton
interactions can be effectively suppressed by selecting events with Large Rapidity Gaps (LRGs) or the leading proton.
Indeed, each additional 'parton-parton$\to$ dijet' scattering needs additional energy and is accompanied by the colour
flow created by the parton cascade needed to form the incoming partons.
This colour flow produces secondaries that fill the LRG. The LRG survival probability, $S^2$, (i.e. the probability not
to destroy the LRG) is rather small, $S^2\leq 0.1$, see e.g. Ref.~\cite{Khoze:2017sdd}\footnote{The value $S^2=0.1$ is
  consistent with the experimental results \cite{CMS:2020dns,ATLAS:2015yqo} for diffractive
high $E_T$ dijet production. Strictly speaking, the gap survival factor
$S^2$ depends on the particular process. However, in our case both the instanton and the high $E_T$ 
dijet are mainly  produced in the collision of two virtual incoming gluons.
Moreover, the virtuality of the gluons producing the QCD instanton 
is expected to be smaller than that for the case of the dijet production,
 corresponding to a somewhat larger impact parameter 
and relatively larger $S^2$. Therefore, our approach
 should be considered rather as a conservative estimate.}. Thus, the probability to
observe $n$ additional branches of parton-parton interactions in LRG events is suppressed by the factor $(S^2)^n$. 

\begin{figure} [t]
\begin{center}
\includegraphics[trim=0.0cm 0cm 0cm 0cm,scale=0.47]{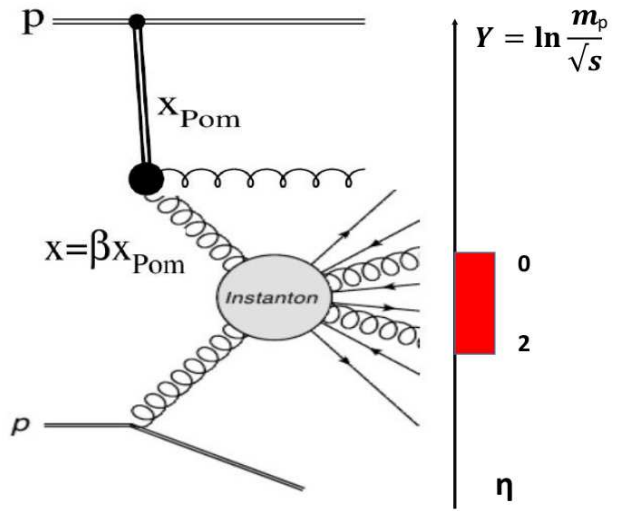}
\caption{\small Instanton production in a diffractive process with an LRG. The Pomeron exchange is shown by the thick doubled line. The red bar shows the range of $\eta$ considered in this paper. $Y$ indicates the incoming proton position in rapidity. As shown in the diagram secondaries will be produced also outside this range but they will not be used when calculating $E_{T}$ or $N_{ch}$.}  
\label{f2}
\end{center}
\end{figure}

For the case of one leading proton, the instanton events are generated as being produced in the proton-Pomeron
collision (see Fig.~\ref{f2} for illustration). When the two leading protons are detected, the Pomeron-Pomeron inelastic
collision is considered (see Ref.~\cite {Khoze:2021pwd} for more details). The incoming Pomeron parton distribution is
taken from the HERA data. In particular, we use the fit B of H1 collaboration~\cite{H1:2006zyl}. The obtained
cross-section is multiplied by the
gap survival probability $S^2=0.1 (0.05)$ for the one (two) leading proton kinematics. To diminish the possible scale
uncertainty we use the $k_T$-factorization approach as it was described in~\cite{Khoze:2021pwd}. The unintegrated PDFs
are calculated based on the LO KMR prescription~\cite{Kimber:2001sc}. We account for the Mueller form
factor~\cite{Mueller:1990qa}, describing quantum corrections to the gluon-gluon collision, and for the incoming gluon
virtuality, $Q^2$, via the corresponding instanton form factor 
\begin{equation}
\label{JQ}
J(\rho Q)~=~\rho QK_1(\rho Q)\ ,
\end{equation}
  where $\rho$ is the instanton radius and $K_1$ is the Bessel function.
 
In this study we work at $\sqrt s = 14$~TeV and we scrutinize two mass regions, namely $M_{\rm inst} > 60$~GeV
and $M_{\rm inst} > 100$~GeV, both
in the FPD acceptance $0.02 < \xi < 0.05$ and for the single-tagged (ST) configuration, the latter also
for the double-tagged (DT) configuration. In total, we then work with three instanton signal event record
samples. All are generated using the RAMBO algorithm \cite{Kleiss:1985gy} at $\xi = 0.03$, with
proton-Pomeron collision type for the ST approach ($2.5\cdot 10^{6}$ events for $M_{\rm inst} > 60$~GeV
and $5\cdot 10^{5}$ events for $M_{\rm inst} > 100$~GeV) and with Pomeron-Pomeron collision type for the
DT approach ($10^{5}$ events). The respective instanton production cross-sections integrated
over the $0.02 < \xi < 0.05$ region are 1004.6~pb and 39.6~pb for the proton-Pomeron
sample and for $M_{\rm inst} > 60$~GeV and 100~GeV, respectively, and 500~fb for the Pomeron-Pomeron
sample and $M_{\rm inst} > 100$~GeV. All samples are showered and hadronised using
PYTHIA~8.2~\cite{Sjostrand:2014zea} with initial-state radiation (ISR), final-state radiation (FSR) and
the MPI switched on. For simplicity, in all cases the instanton signal is generated with a forward leading
proton only in one hemisphere with $p_z < 0$. For this reason, in the following all studies are done and
all cuts are tailored for one hemisphere. Assuming a full symmetry in $z$ coordinate, final numbers are
then obtained by doubling those from the studied hemisphere.

For both, ST and DT approaches, we consider two backgrounds, dijet production in Single-diffractive (SD) and
Non-diffractive (ND) interactions, which are by far most dominant due to resemblance of their final states to
the final state of the
signal and due to their huge production cross-sections with respect to that of the signal. In both, again ISR, FSR
and MPI are switched on. For the SD dijet background, we use the dynamical gap survival approach
with MPI between Pomeron and proton switched on~\cite{Rasmussen:2015qgr}. Production cross-sections for
$\hat{p}_{\rm T,min} > 10$~GeV are 80~$\mu$b for SD and 8.64~mb for ND. At truth level, we have generated
roughly $5\cdot 10^{11}$ SD events and $8.5\cdot 10^{11}$ ND events.

Since the relatively heavy instanton produces  a rather large number of jets with the energy $E_T\sim 1/\rho$ we are
looking for high multiplicity events which:
\begin{itemize}
\item do not contain very high-$E_T$ jets, and 
\item still have a large density of the transverse energy, $\sum_i dE_{Ti}/d\eta\sim M_{\rm inst}/3$ (the sum is over all  secondary particles in the given $\eta$ interval).
\end{itemize}

\medskip 
Moreover, since each jet from the instanton cluster contains a leading hadron 
we can select events with a large multiplicity (say, $N_{\rm ch} >20$) of charged particles with $p_T>0.5$~GeV
in a limited rapidity interval.  
Since the LHC detectors never cover the whole ($4\pi$) rapidity interval, there is no chance to adequately measure the
value of $M_{\rm inst}$. To select the events with appropriate $M_{\rm inst}$ we introduce the cut on the total transverse
energy measured within the given rapidity interval $\sum_i E_{Ti}>M_0$.
  
Here we consider the instanton production in the proton-Pomeron collision (in terms of Regge theory  the Pomeron
exchange is responsible for the presence of the LRG) selecting  events with a large multiplicity and relatively large
transverse energy~\footnote{The idea to search for the instanton in events with very large multiplicity but not too
large transverse energy and the first evaluation of the instanton cross-section at collider energies were discussed
long ago in Ref.~\cite{Balitsky:1993sg}.}.
We expect that these events will be more or less spherically symmetric, that is, in such events there should be a large probability to observe the sphericity $S$ close to 1. Since we can not observe the particles in the whole $4\pi$ sphere we consider the ``transverse sphericity" (or  cylindricity) defined as $S_T=2\lambda_2/(\lambda_1+\lambda_2)$ where $\lambda_i$ are the eigenvalues of the matrix
  \begin{equation}
  \label{st1}
  S^{\alpha\beta}=\frac{\sum_i p^\alpha_i p^\beta_i}{\sum_i \lvert \vec p^2_i\rvert}\ ,
  \end{equation}
  and $\lambda_2<\lambda_1$.
Here $p_i^\alpha$ is the two-dimensional transverse component of the momentum of the $ i$-the particle and we sum over all particles observed in the event within a given rapidity interval.\\

\section{Search strategy}\label{sec:strategy}
The search for instanton signal in the harsh environment of various backgrounds consists of two steps. In the first
step, we work at generator level, examine several cut scenarios and select the one giving the best signal to background
(S/B) ratio, which we then call a ``golden scenario''. In the second step, a fast simulation of the detector and
pile-up effects is added, and their impact on the S/B ratio for this golden scenario is investigated.

As reported in publications by ATLAS and CMS where forward proton
detectors AFP and CT-PPS, respectively, were used, the lowest $\xi$ value reachable in Run~2 was 0.02
and a similar reach is expected for Run~3.
Regions of higher $\xi$ may be contaminated by Reggeon contributions and ND events that survive the
$\xi>0.02$ cut thanks to fluctuations in the hadronization process and the large cross-section. Therefore,
to suppress both these contributions, we decided to work in a rather narrow $0.02 < \xi < 0.05$ range
which is used as a baseline acceptance region in all considerations below. We study two approaches,
using single-tagged and double-tagged protons. While in the former, the selection efficiency is higher but
background contamination is usually higher and one cannot use the ToF detector to suppress pile-up
background, in the latter, all backgrounds are usually better suppressed including the pile-up by utilizing
the ToF detector but we pay for that by lower selection efficiencies. 
Nevertheless in the end it depends on production cross-sections of signal and backgrounds
and on the impact of various cut scenarios for a chosen FPD acceptance and a given luminosity scenario.  
Since pile-up effects can easily diminish advantages of the final
state of the signal (including the presence of LRG), we concentrate on low pile-up scenarios in the
case of single-tagged approach and on medium pile-up amounts in the double-tagged approach.

\subsection{Combinatorial background}\label{sec:comb}
With the increasing average number of pile-up events per bunch crossing, $\langle\mu\rangle$, the probability
to detect a pile-up proton in the FPD acceptance increases. When overlaid with a hard-scale event (triggered
by various L1 triggers) it forms a combinatorial background which is usually dangerous due to resemblance
of its final state to that of the signal and due to the fact that both, the pile-up proton seen in FPD
and the hard-scale event can occur much more frequently than the signal. Detailed discussions and dependences
on $\langle\mu\rangle$ of this probability to fake either single-tagged (ST) or double-tagged (DT) signal
in FPDs, and how this background can be tamed by using time-of-flight (ToF) detectors installed in FPDs at
LHC can be followed in Refs.~\cite{Cerny:2020rvp,Tasevsky:2014cpa,Harland-Lang:2018hmi}. Frequency of both,
the fake ST and DT signal in FPD, depends on the probability to see a proton from minimum bias events
in the acceptance of FPD on one side, $P_{\rm ST}$. For the acceptance considered here, namely
$0.02 < \xi < 0.05$, $P_{\rm ST} = 0.0048$ as predicted by Pythia~8.2 at $\sqrt{s} = 14$~TeV for minimum bias events
with ISR, FSR and MPI switched on. For a given amount of pile-up, $\langle\mu\rangle$,
the probability per bunch crossing to have a pile-up proton in the acceptance of FPD on one side is

\begin{equation}
  P_{\rm comb} = 1 - (1 - P_{\rm ST})^{\langle\mu\rangle} .
\label{comb}
\end{equation}

This combinatorial factor serves to estimate the total combinatorial background for both, the ST and DT strategies.
The most dangerous situation for ST occurs when one hard-scale event (for example SD or ND dijets with low jet
$p_T$) is overlayed with another soft SD event producing a forward going proton. For the DT strategy, the danger lies
in an overlay of three events: one hard-scale SD or ND dijet event and two soft SD events each giving a forward
going proton on opposite sides from the IP. The total combinatorial background is then estimated as a product of the
fiducial cross section for the SD or ND dijet process (i.e. after applying all cuts except the $\xi$ acceptance) times
the combinatorial factor $P_{\rm comb}$ in the ST case or $P_{\rm comb}^2$ in the DT case. We remind that the $\xi$
acceptance cut is applied already on the one soft SD event for ST (two soft SD events for DT), so it cannot be applied
anymore on the hard-scale event. 



The values of the combinatorial factor are then 0.48\%, 0.96\% and 2.38\% for $\langle\mu\rangle$ of 1, 2
and 5, considered for ST, and 0.84\% and 4.58\% for $\langle\mu\rangle$ of 20 and 50, considered for DT. 
The last two can be further reduced by making use of ToF
detectors (we need to detect a proton on each side in one event). If we assume a resolution of 10~ps,
the ToF suppression of the combinatorial background by a factor of about 18 (16) for $\langle\mu\rangle$
of 20 and 50, respectively, can be achieved.
We also note that production cross
section of the ST (DT) instanton signal is roughly 100 (1000) times smaller than the corresponding
instanton signal cross-section in inclusive events. The reasoning is based on the presence of $S^2$ in the
case of diffractively produced instanton and on the ratio of diffractive to inclusive PDFs being about 10.

\subsection{Generator level}
In addition to the FPD acceptance, the following variables based on properties of particles measured in
the central detector were examined (note that we are limiting ourselves to positive pseudorapidities
because the leading protons in signal samples are generated only with negative rapidities):
\begin{itemize}
\item $N_{\rm ch05}$ = number of charged particles with $p_T > 0.5$~GeV and detected in a part of the
  central tracker $0.0 < \eta < 2.0$
\item $N_{\rm ch20(25,30)}$ = number of charged particles with $0 < \eta < 2.0$ and $p_T > 2.0, 2.5$ or 3.0~GeV
\item $\sum E_T$ = sum of $E_T$ of charged particles with $p_T > 0.5$~GeV and $0.0 < \eta < 2.0$
\item $N_{\rm ch05fw}$ = number of charged particles with $p_T > 0.5$~GeV detected in one half of the
  forward calorimeter $2.5 < \eta < 4.9$
\item $\sum E_T^{\rm fw}$ = sum of $E_T$ of charged particles with $p_T > 0.5$~GeV and $2.5 < \eta < 4.9$  
\end{itemize}  

\begin{figure*}[h]
\begin{center}  
  \includegraphics[width=0.49\textwidth,height=5cm]{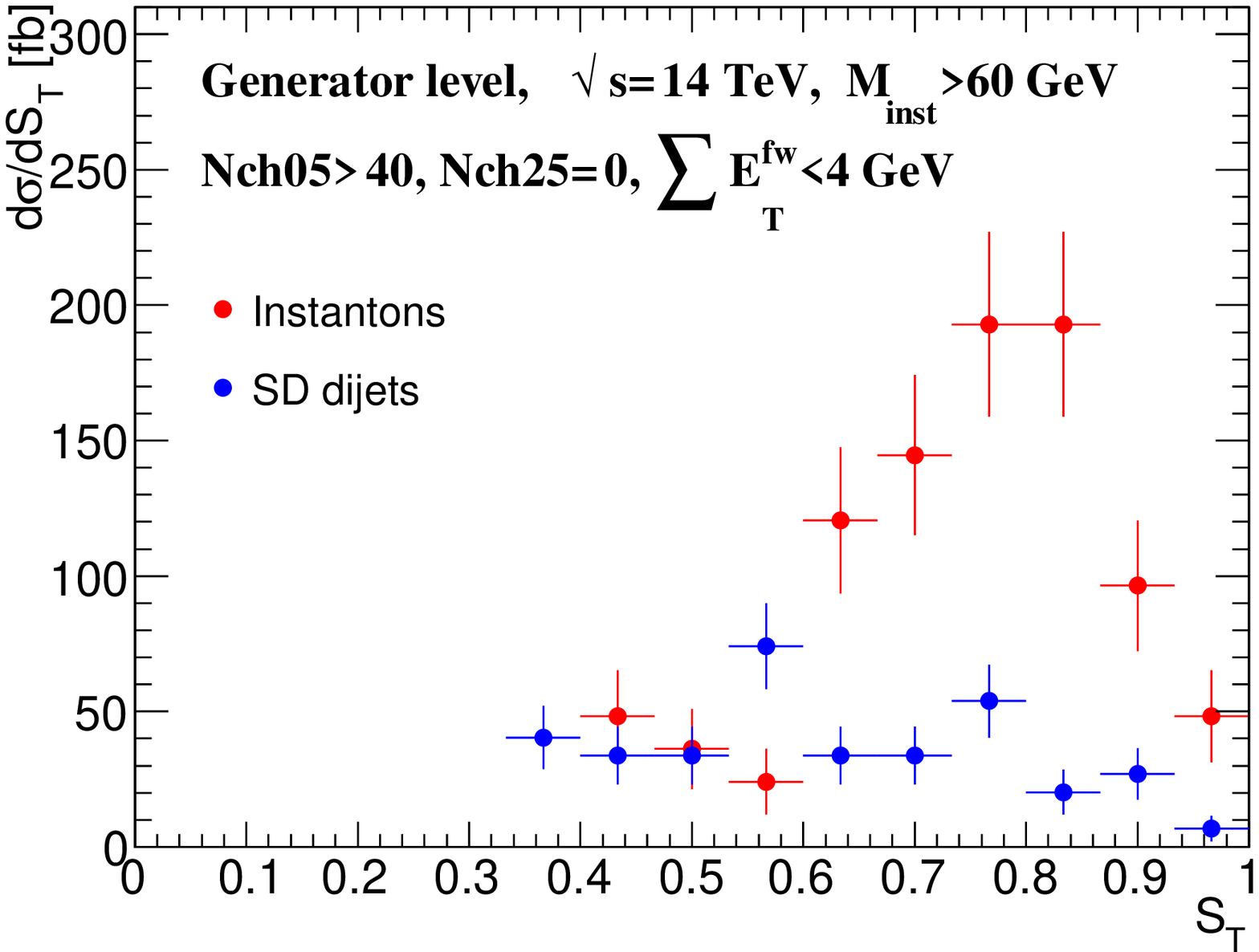}
\includegraphics[width=0.49\textwidth,height=5cm]{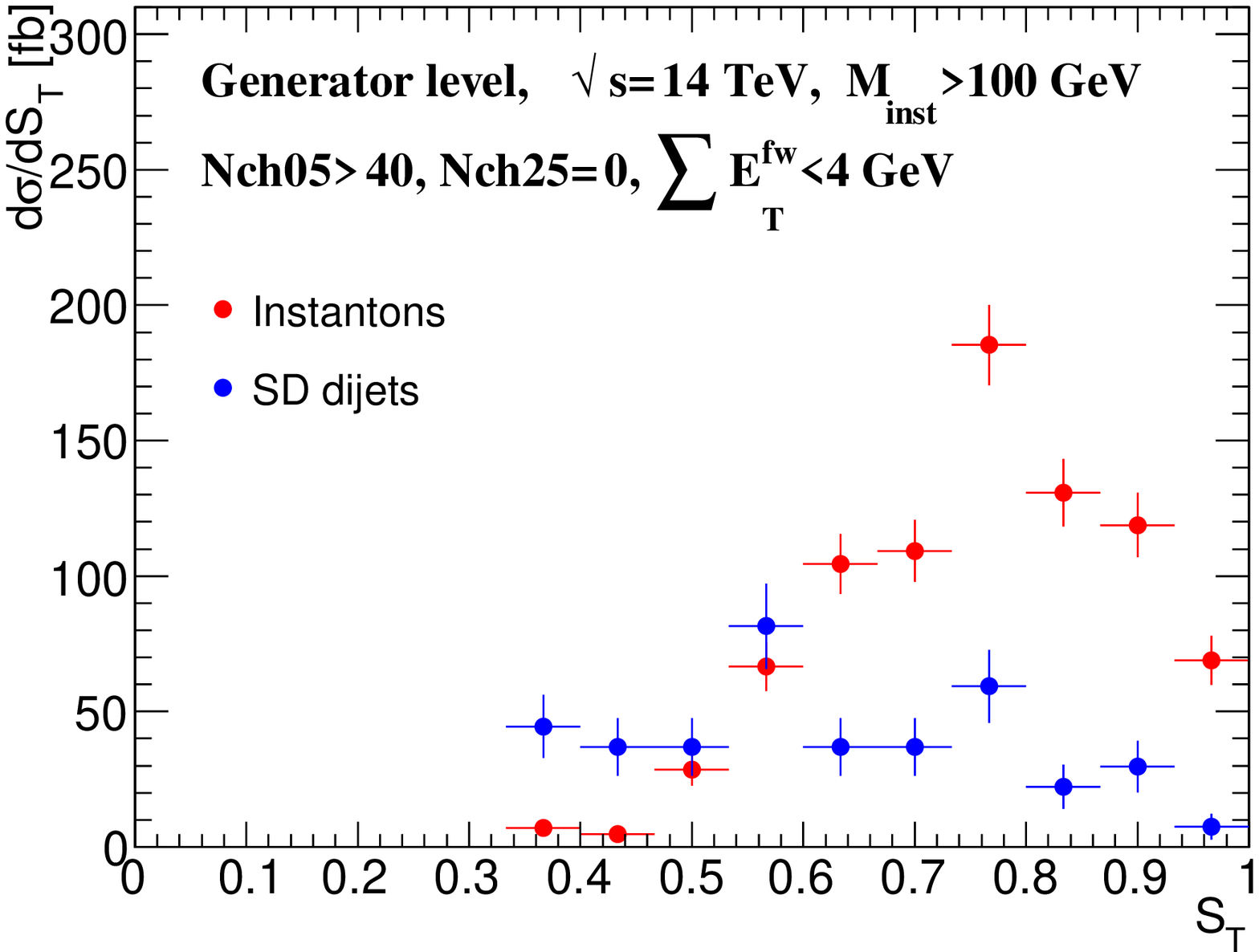}  
\caption{Differential cross-section as a function of transverse sphericity at generator level
  separately for the instanton signal from proton-Pomeron collisions ($M_{\rm inst} > 60$~GeV on the left and
  $M_{\rm inst} > 100$~GeV on the right) generated by RAMBO and SD dijet background generated by PYTHIA~8.2 for the
  golden cut scenario (Eq.~(\ref{Eq:goldenscenario})). Only statistical uncertainties are plotted.}
\label{fig:stgen}
\end{center}
\end{figure*}

From each variable we constructed at least one cut as follows: $N_{\rm ch05} > 30(40)$, $N_{\rm ch20(25,30)} = 0$,
$\sum E_T > 30(40)$~GeV, $N_{\rm ch05fw} < 6$ and $\sum E_T^{\rm fw} < 4$~GeV. While vetoing particles with
a relatively high transverse momentum aims at rejecting events with high-$E_T$ jets, the cuts on the
activity in the forward calorimeter are introduced in order to suppress a fake signal caused by the MPI. 
In total we have examined 42~cut scenarios on the proton-Pomeron sample with $M_{\rm inst} > 100$~GeV. From comparing
fiducial cross-sections of signal and background for each cut scenario using the signal sample and a very
large SD sample, we conclude that the best S/B ratio of 2.1 is obtained for two cut scenarios, namely one
with

\begin{equation}
N_{\rm ch05} > 40 {\rm \ \ and \ \ } N_{\rm ch25} = 0 {\rm \ \ and \ \ } \sum E_T^{\rm fw} < 4~{\rm GeV}
  \label{Eq:goldenscenario}
\end{equation}

and one with the same cuts as in Eq.(\ref{Eq:goldenscenario}) but with $\sum E_T > 30$~GeV in addition.
Since the ratio S/B after adding the cut $\sum E_T > 30$~GeV remains the same, we drop this second one and
call the cuts in Eq.(\ref{Eq:goldenscenario}) the ``golden cut scenario''. For this scenario the
corresponding S/B for the signal sample with $M_{\rm inst} > 60$~GeV is 2.3. When applying the same set
of cuts on a large sample of ND dijet events (over 8.5$\cdot 10^{11}$ events were generated), none
survives. This translates to an estimated fiducial cross-section for ND background to be smaller
than 10.2~fb. The differential cross-section as a function of transverse sphericity, $S_T$, for the signal
and SD dijet background and for the golden cut scenario is shown in Fig.~\ref{fig:stgen}. For both
proton-Pomeron signal samples, $M_{\rm inst} > 60$~GeV and $M_{\rm inst} > 100$~GeV, the signal is clearly seen to be
well-separated from the SD background and peaking at higher $S_T$ values as expected. By comparing
the left with the right plot, we can say that the cuts from the golden scenario reduce to a large extent
the contribution from $M_{\rm inst} < 100$~GeV.

\subsection{Detector level}\label{SDdet}
The detector and pile-up effects for events surviving the golden cut scenario are then studied using
Delphes~3.5 fast simulation package~\cite{Forthomme:2018ecc} with an ATLAS input card.
On both, the signal and background samples of events at generator level, we run Delphes with
various amounts of pile-up, defined by an average number of pile-up events per bunch crossing,
$\langle\mu\rangle$. For each pile-up amount, we are also considering an adequate integrated
luminosity, which leads us to the following working ($\langle\mu\rangle$,$\cal L$ [fb$^{-1}$]) points:
(0,0.1), (1,0.1), (2,1) and (5,10) for the ST approach and (20,60) and (50,300) for the DT approach.
Given the assumed difficulties to distinguish the signal from all relevant backgrounds including pile-up,
we rather concentrate on low (in ST approach) and medium (in DT approach) amounts of pile-up. At the
same time, we are convinced that these are conceivable luminosity scenarios for Run~2 and Run~3.

At detector level, we work with tracks in the central tracker and with calorimeter clusters in the
forward calorimeter. On the sample of signal events, we first need to tune track selection
criteria and cuts that are based on tracks. Following the track selection procedure used in the
analysis of charged tracks at 13~TeV by ATLAS~\cite{ATLAS:2016zba}, we reduce tracks from overlaid pile-up
events to an acceptable minimum. Although we observe more than twice as many tracks with $p_T > 0.5$~GeV
for $\langle\mu\rangle = 5$ than for no pile-up, after requiring $(z_{\rm vtx}-z_{\rm trk})\sin\theta < 1.5$~mm
and $d_0^{\rm trk} < 1.5$~mm in addition, the number of tracks with $p_T > 0.5$~GeV drops considerably and
exceeds the one at zero pile-up by only 2.5\%. Here
$z_{\rm vtx}$ is the z-coordinate of the primary vertex, $z_{\rm trk}, \theta$ and $d_0^{\rm trk}$ are
z-coordinate, polar angle and transverse impact parameter of a given track, respectively.
Track reconstruction efficiencies and resolutions as functions of track $p_T$ and $\eta$ are accounted
for via Delphes input card. A nominal track efficiency is roughly 80\% on average for $p_T > 0.5$~GeV
and not close to the tracker edges (see e.g. Ref.~\cite{ATLAS:2016vxz}, Fig.~13) and this one is applied
in all studies. As a systematic cross-check, we also examine an optimized track efficiency 
(see Ref.~\cite{Schillaci:2021zcr}) which reaches almost 90\% on average for $p_T > 0.5$~GeV.
Our aim is to illustrate the situation after data taking, thus
to estimate real numbers of collected events, therefore we do not correct for these track reconstruction
inefficiencies but rather adapt the first cut in Eq.(\ref{Eq:goldenscenario}) to $N_{\rm tr05} > X$
(where $N_{\rm tr05}$ is number of tracks with $p_T > 0.5$~GeV and $0 < \eta < 2.0$ and $X < 40$).
The energy flow measurement from Run~1 by ATLAS~\cite{ATLAS:2012vre} indicates that the third
cut in Eq.(\ref{Eq:goldenscenario}) needs to be adapted as well. As Fig.~2 a in Ref.~\cite{ATLAS:2012vre}
shows, Pythia~6 predicts a steeper distribution of $\sum E_T$ in the forward region $4.0 < \lvert\eta\rvert < 4.8$
than data, both at detector level. While for $\sum E_T < $~3--4~GeV, it overestimates the data, for 
$\sum E_T > $~3--4~GeV it underestimates them. To account for this Pythia~6 discrepancy, one would need
to re-weight it by a ratio data to MC at detector level as is done in Ref.~\cite{ATLAS:2012vre}.
But taking into account two small caveats, namely that in our analysis we work with Pythia~8.2 and
in the larger region, $2.5 < \eta < 4.9$, we decide to take rather a simplistic approach and adapt
the third cut in Eq.(\ref{Eq:goldenscenario}) to $\sum E_T^{\rm fwcalo} < 5$ (or 6)~GeV (where the sum runs
over clusters in the forward calorimeter). At the same time, this new threshold should not be too
far from the original 4~GeV threshold used at generator level. The reason is that clusters in the
forward calorimeter lie outside the tracker coverage, so they cannot be linked with tracks
and consequently we do not have information about which part of cluster energy comes from pile-up. 
Therefore, if departing too far from the 4~GeV threshold, we would be picking up an uncontrollable
amount of pile-up. Furthermore, we should also take into account that detector and pile-up effects may
cause migrations of events from regions below to regions above cut thresholds (for example at generator
level, $N_{\rm ch05} < 40$ but pile-up events may cause $N_{\rm tr05} > 40$).
To this end, out of many generated ND and SD background events ($2\cdot 10^{11}$ for SD, $6\cdot 10^{11}$
for ND) we retain subsamples of events surviving cuts that are more relaxed than those in Eq.(\ref{Eq:goldenscenario}),
namely we apply cuts $N_{\rm ch05} > 30$ and $N_{\rm ch30} = 0$ and $\sum E_T^{\rm fw} < 6$~GeV. Those
samples (about 100 (1500) thousand of SD (ND) events) are then processed with Delphes and subsequently
four samples with different amounts of pile-up events are created for each instanton mass region
and analyzed in detail.
For the sake of completeness, we note that to save computational power, the proton $\xi$ cut
has not been applied even for the zero pile-up scenario and its effect has been accounted for by scaling the
fiducial cross section for SD (ND) dijets by $P_{\rm ST}(\mu=0)$ = 4.3\% (0.015\%). These $P_{\rm ST}$ values correspond
to SD (ND) dijet backgrounds with $\hat{p_{\rm T}} >$ 10~GeV. On another large-statistics SD dijet sample we have
applied the proton $\xi$ acceptance cut and got very similar results as using the SD dijet sample without this
preselection. This confirms the validity of this factorization of the FPD $\xi$ acceptance cut.

As explained above, to avoid double-counting, the generator-level FPD $\xi$-acceptance cut obtained from the proton
information cannot be applied for the hard-scale event since it is already applied on overlayed soft pile-up events.
But we can make use of information from the central detector and try to restrict ourselves to a $\xi$-region similar
to that of the applied FPD acceptance $0.02 < \xi < 0.05$ at generator level - note that a strict matching is not
possible using tracker information only due to its too narrow $\eta$-acceptance but the $\xi$ quantity evaluated using
the calorimeter (whose coverage is $\mid\eta\mid<4.9$), $\xi^{\rm calo}$, can bring us much closer to the generator-level
$\xi$-range. The additional cut on $\xi^{\rm calo} < 0.025$ turns out to be reasonably efficient. Following the
discussion above, we adapt the golden cut scenario in Eq.(\ref{Eq:goldenscenario}) to these two sets of cuts at
detector level:

\begin{equation}
  N_{\rm tr05} > 25 {\rm \ \ and \ \ } N_{\rm tr20} = 0 {\rm \ \ and \ \ } \sum E_T^{\rm fwcalo} < 5~{\rm GeV \ \ and \ \ }
  \xi^{\rm calo} < 0.025
  \label{Eq:goldendet60}
\end{equation}

\begin{equation}
N_{\rm tr05} > 30 {\rm \ \ and \ \ } N_{\rm tr25} = 0 {\rm \ \ and \ \ } \sum E_T^{\rm fwcalo} < 5~{\rm GeV \ \ and \ \ }
  \xi^{\rm calo} < 0.025
  \label{Eq:goldendet100}
\end{equation}

where $N_{\rm tr05}$, $N_{\rm tr25}$ and $N_{\rm tr20}$ are numbers of tracks in the region $0.0 < \eta < 2.0$
and for $p_T > 0.5$~GeV, $p_T > 2.5$~GeV and  $p_T > 2.0$~GeV, respectively, and $E_T^{\rm fwcalo}$ is a sum
of $E_T$ of clusters in the forward calorimeter with $p_T > 0.5$~GeV and $2.5 < \eta < 4.9$. The $\xi^{\rm calo}$
quantity is calculated as a sum of $E_{\rm T}e^{-\eta}$ over calorimeter clusters with $E_{\rm T} > 0.2$~GeV. 
The cuts in Eq.~(\ref{Eq:goldendet60}) (Eq.~(\ref{Eq:goldendet100})) are used as nominal for
$M_{\rm inst} > 60$~GeV ($M_{\rm inst} > 100$~GeV).

\section{Results}\label{sec:results}
\subsection{Single tag}
\begin{figure*}
  \includegraphics[width=0.5\textwidth,height=5cm]{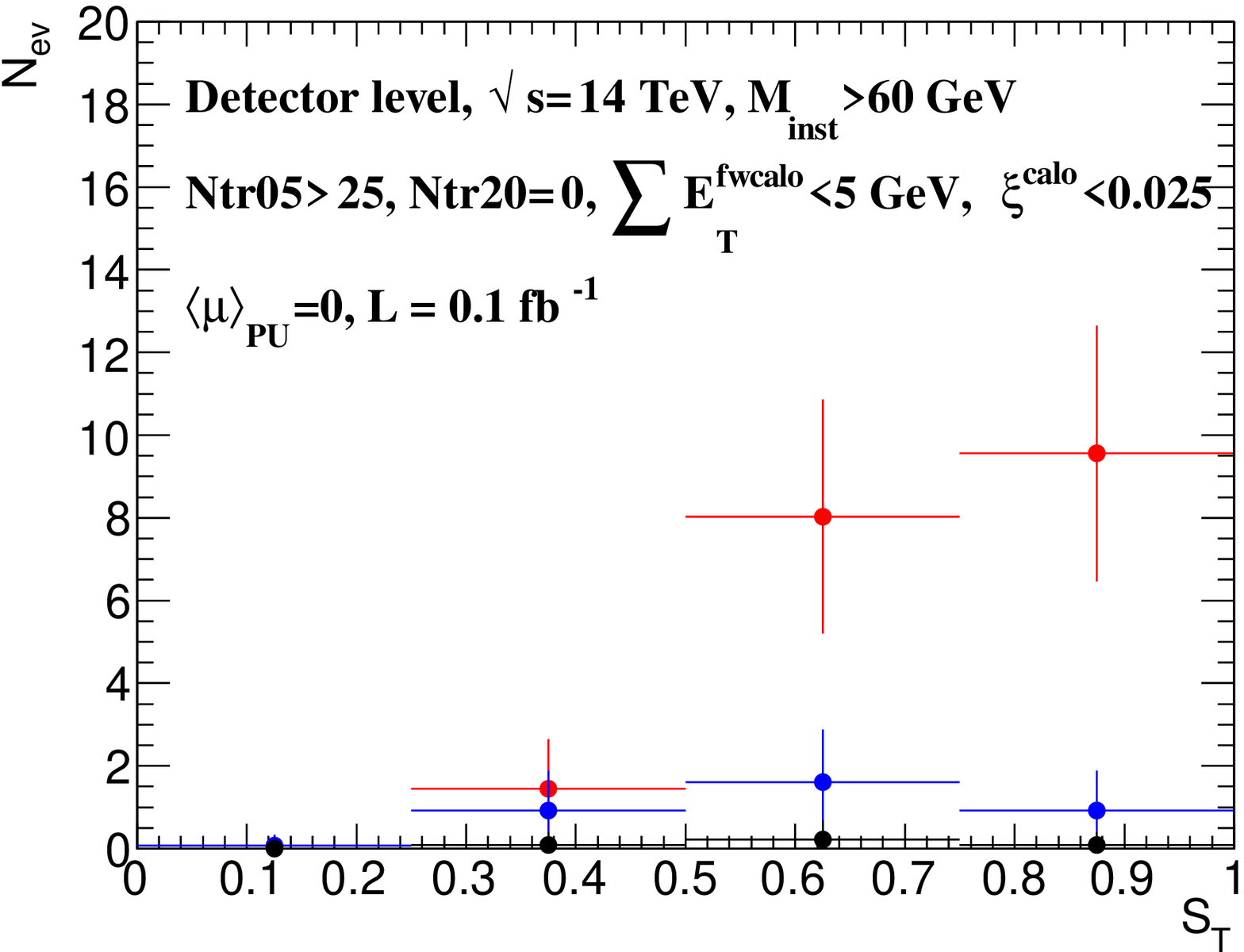}
  \includegraphics[width=0.5\textwidth,height=5cm]{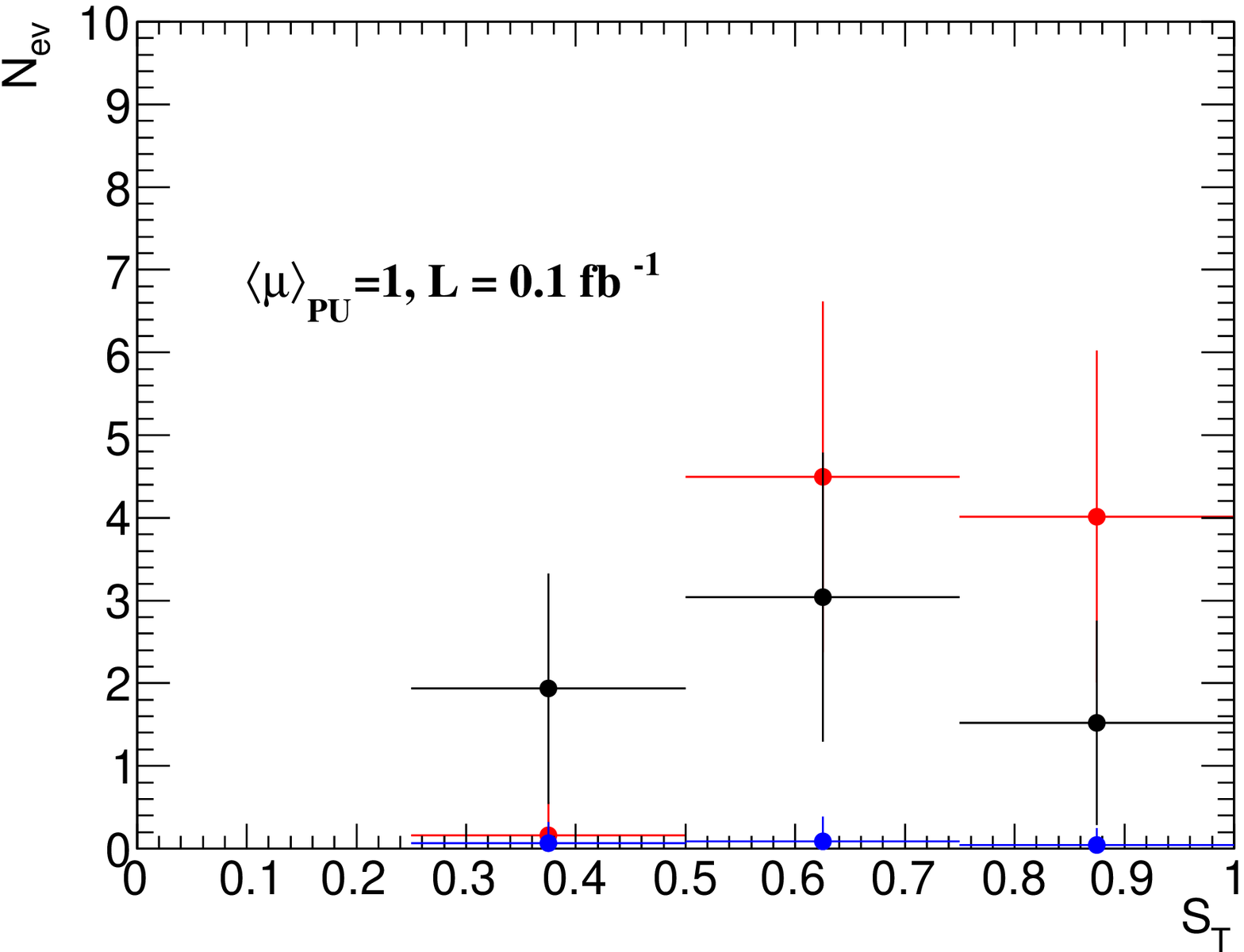}
  \includegraphics[width=0.5\textwidth,height=5cm]{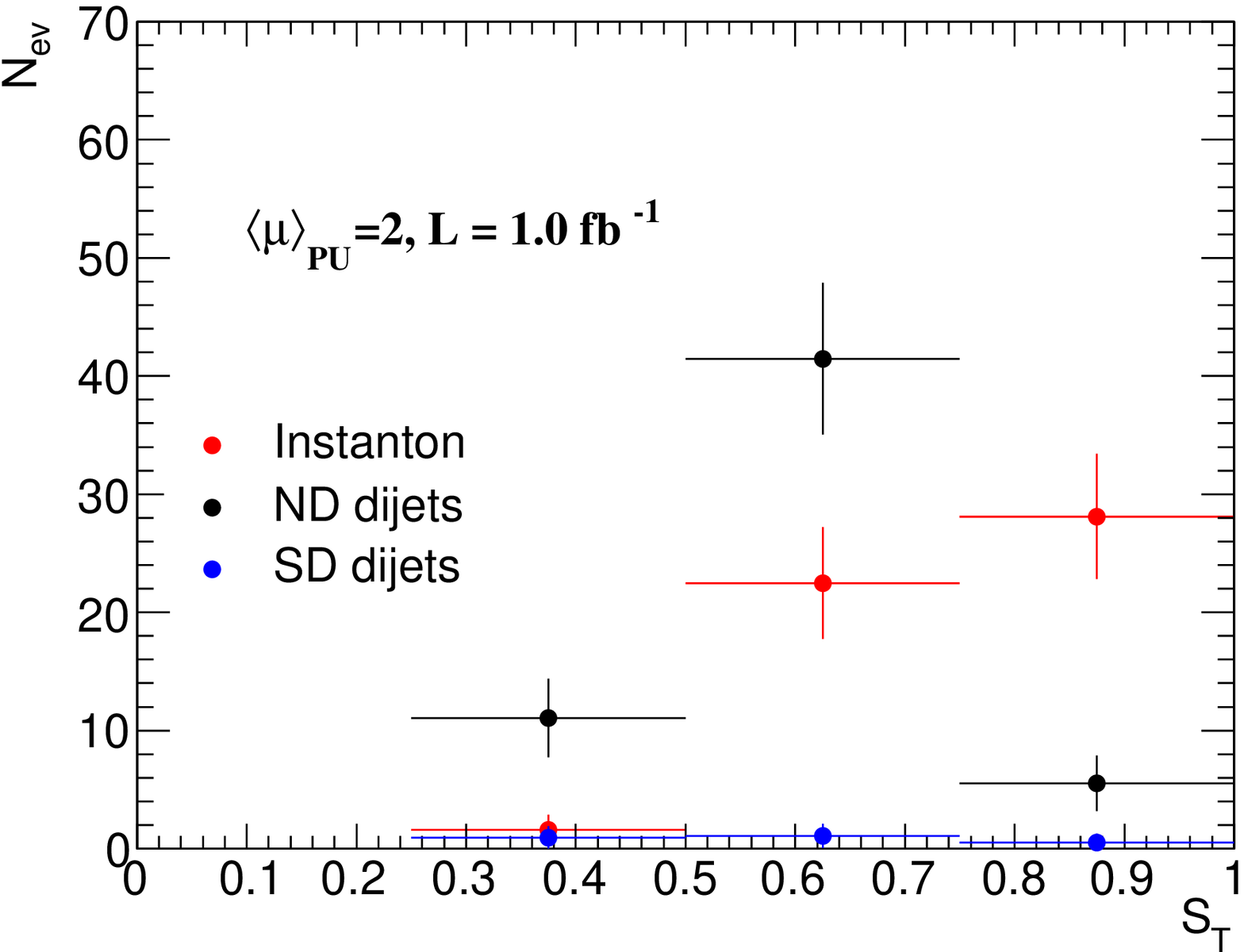}
  \includegraphics[width=0.5\textwidth,height=5cm]{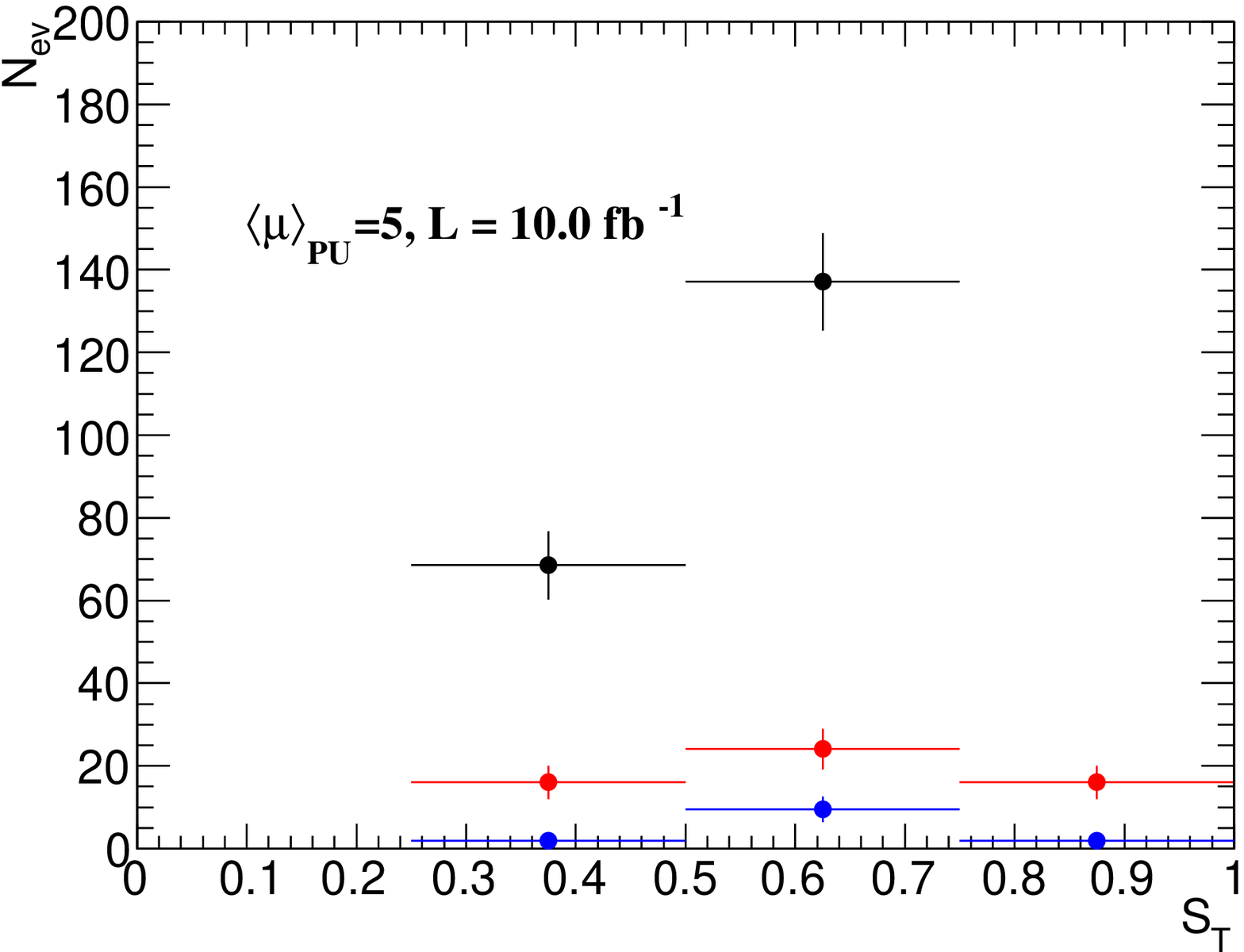}  
  \caption{Distributions of expected event yields as functions of transverse sphericity at detector
    level for instanton signal from proton-Pomeron collisions generated by RAMBO for $M_{\rm inst} > 60$~GeV and
    backgrounds from ND dijets and SD dijets generated by PYTHIA~8.2 after applying detector-level cuts in
    Eq.~(\ref{Eq:goldendet60}) for four luminosity scenarios. Only statistical uncertainties are shown, estimated
    using expected event numbers from Table~\ref{tab:sig}.}
\label{fig:stdet60}
\end{figure*}

\begin{figure*}
  \includegraphics[width=0.5\textwidth,height=5cm]{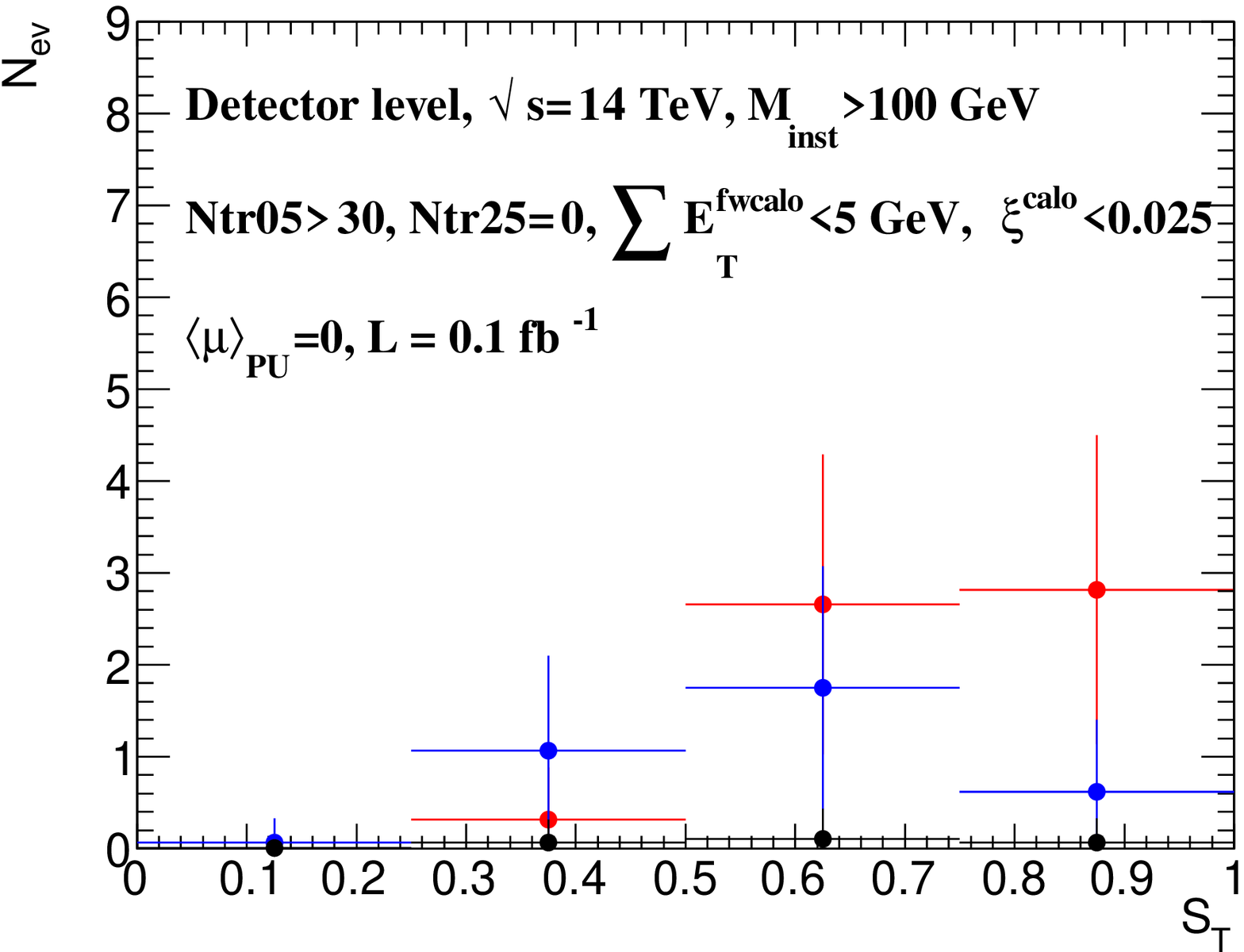}
  \includegraphics[width=0.5\textwidth,height=5cm]{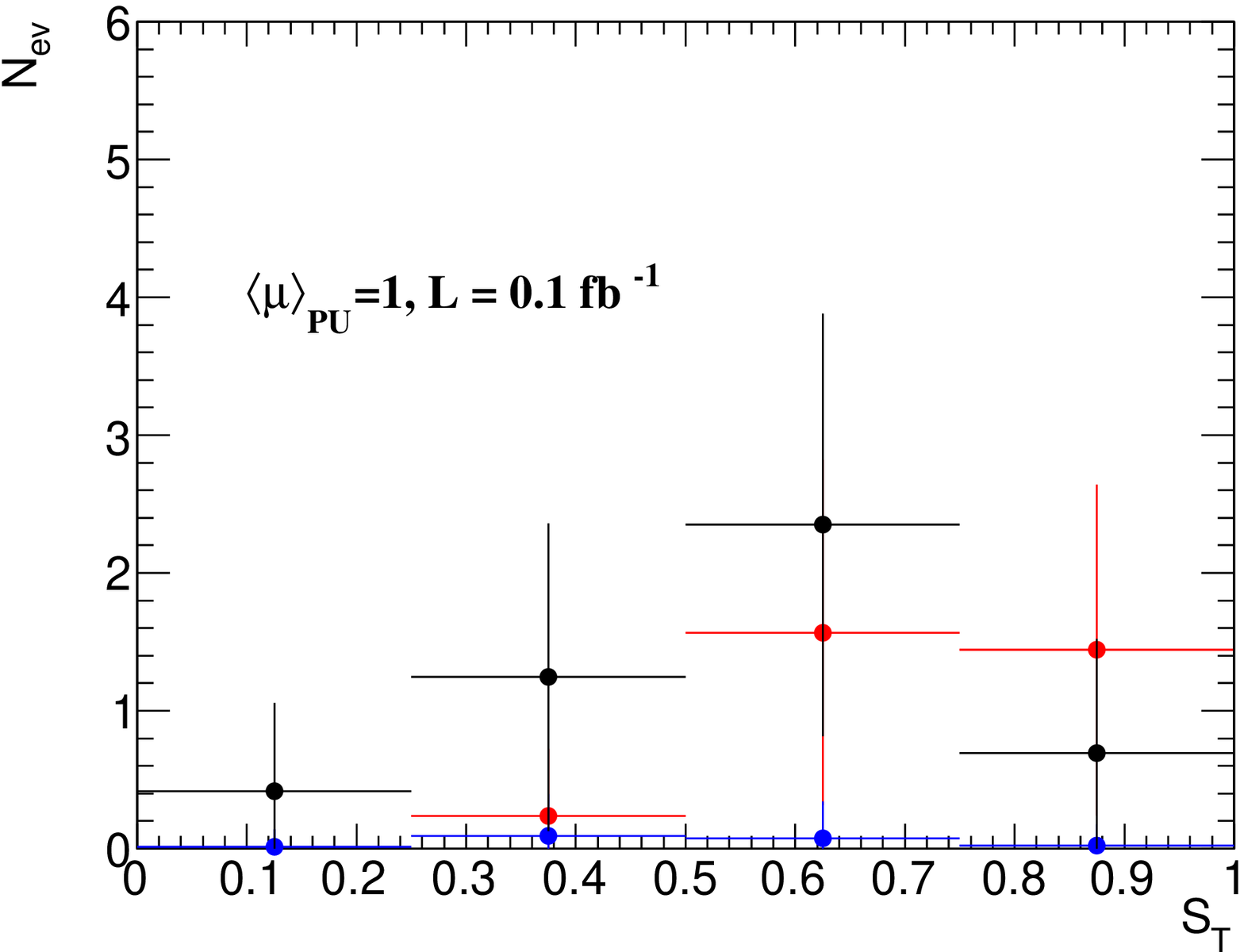}
  \includegraphics[width=0.5\textwidth,height=5cm]{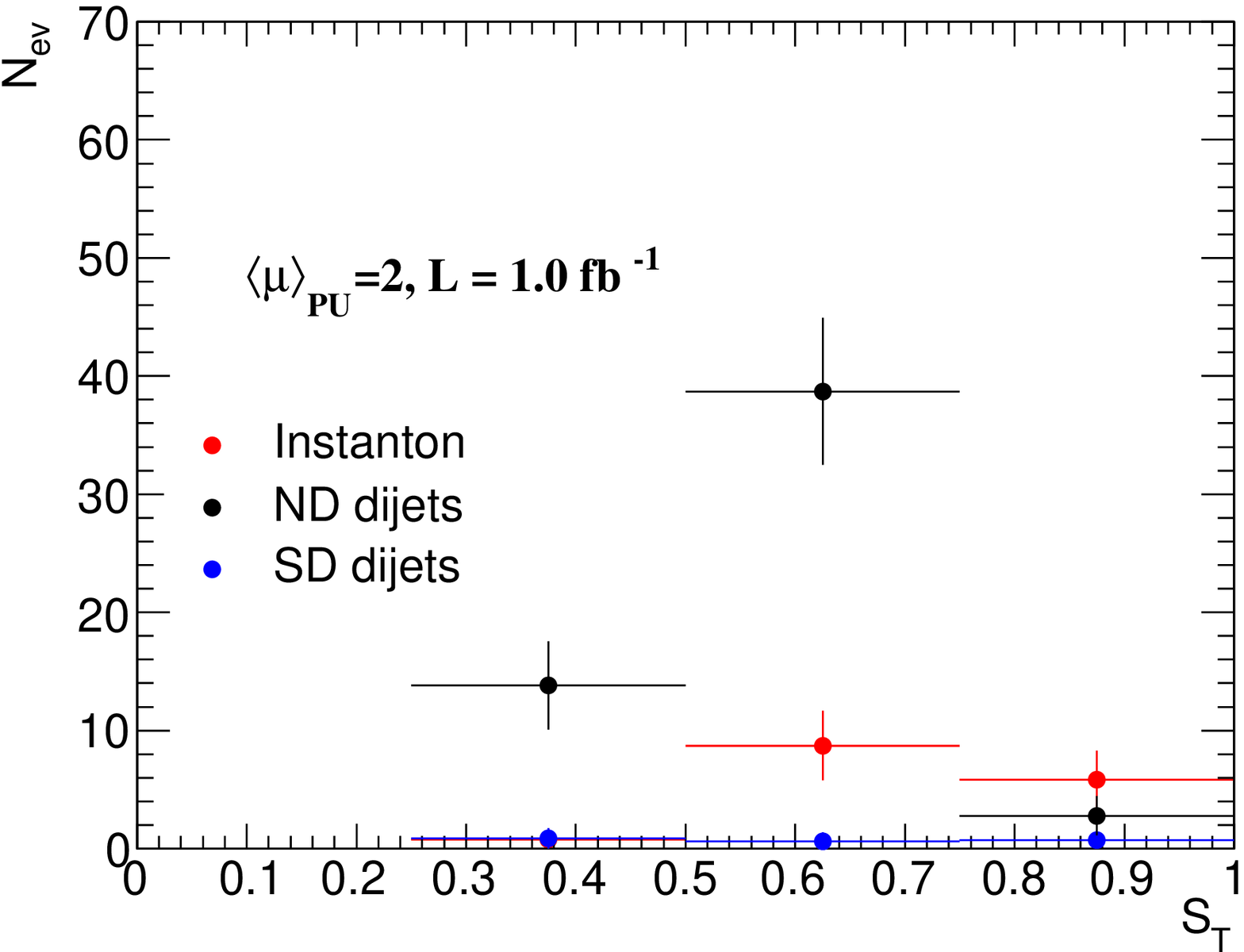}
  \includegraphics[width=0.5\textwidth,height=5cm]{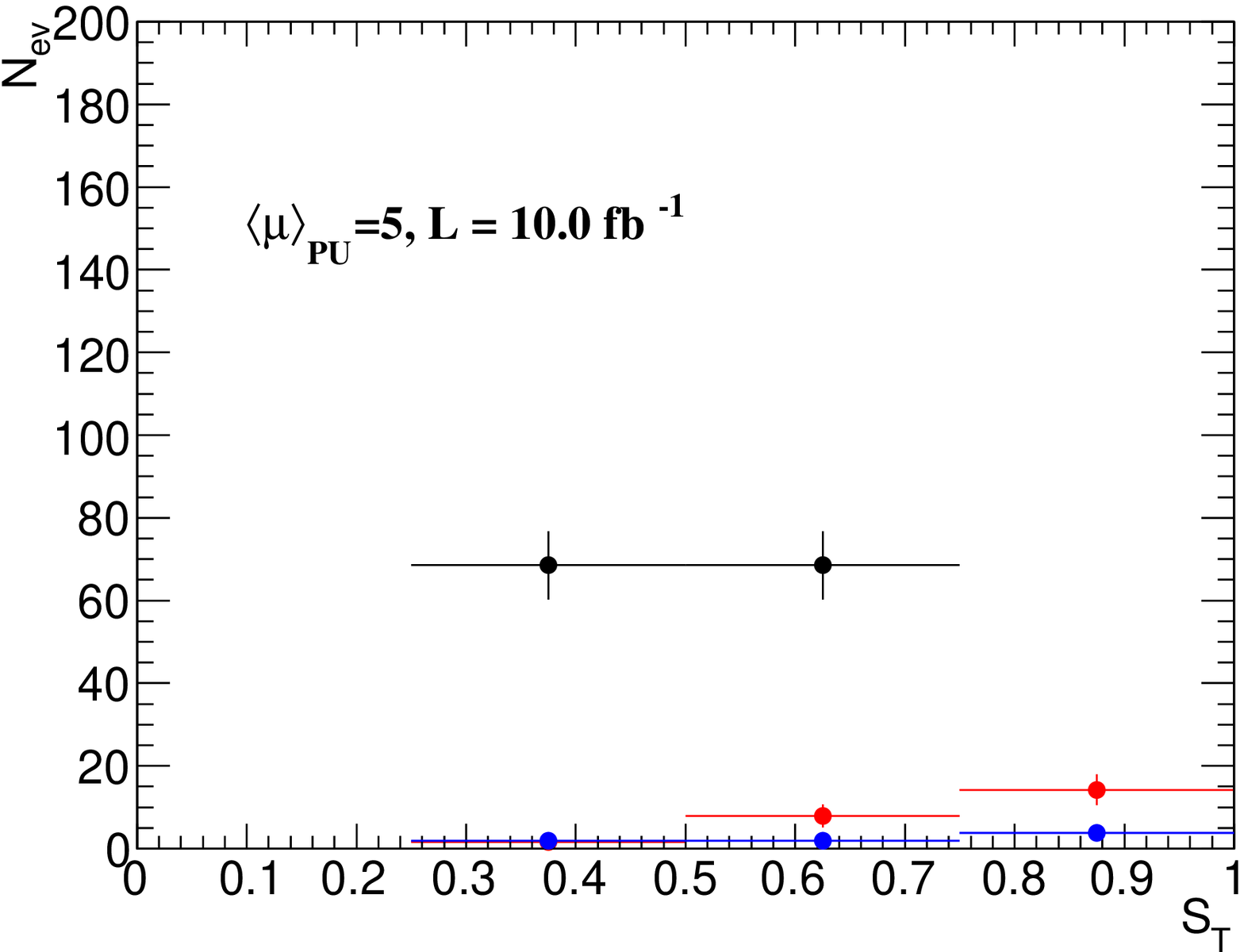}  
  \caption{Distributions of expected event yields as functions of transverse sphericity at detector
    level for instanton signal from proton-Pomeron collisions generated by RAMBO for $M_{\rm inst} > 100$~GeV and
    backgrounds from ND dijets and SD dijets generated by PYTHIA~8.2 after applying detector-level cuts in
    Eq.~(\ref{Eq:goldendet100}) for four luminosity scenarios. Only statistical uncertainties are shown, estimated
    using expected event numbers from Table~\ref{tab:sig}.}
\label{fig:stdet100}
\end{figure*}

To illustrate the situation after data taking at detector level, the expected event yields for signal
generated with $M_{\rm inst} > 60$~GeV and $M_{\rm inst} > 100$~GeV together with both backgrounds, the SD dijets and ND
dijets, are shown in Figs.~\ref{fig:stdet60} and \ref{fig:stdet100} as functions of $S_T$ for four luminosity scenarios,
defined above after applying detector-level cuts defined in Eq.(\ref{Eq:goldendet60}) and Eq.(\ref{Eq:goldendet100}),
respectively.

To estimate the influence of detector effects, we remove the $\xi^{\rm calo} < 0.025$ cut and compare the S/B
    ratio at zero pile-up with the generator-level S/B ratio. While the latter amounts to 2.3 (2.1) for
    $M_{\rm inst} > 60$ (100)~GeV, the former drops to about 0.6 (0.3) which suggests that the detector
effects are noticeable. To study their origin in more detail, we investigate event samples without the
$\xi^{\rm calo} < 0.025$ cut in the following.
For the $M_{\rm inst} > 60$~GeV signal, the mean values of $\sum E_T$ in the forward region for these two cases
 (9.2~ GeV vs. 10.1~GeV) do not differ much hence the track reconstruction inefficiencies and resolutions
 seem to be responsible for the observed S/B deterioration. By adding more and more pile-up, the mean
 values of $\sum E_T^{\rm fwcalo}$ in signal increase to 17, 24 and 37~GeV for $\langle \mu \rangle$ of
 1, 2 and 5, respectively, giving rise to fewer and fewer events surviving the $\sum E_T^{\rm fwcalo}$
 cut with increasing
 pile-up. But the same holds for the SD background, where corresponding values are 3.5 and 4.1~GeV
 for generator level and detector level without pile-up, which then rise with increasing pile-up to
 11.5, 18.7 and 41~GeV. The lower mean values for the SD background are due to the
 pre-selection procedure applied to the SD sample. We remind that we pre-select SD events at generator
 level by cuts defined in Section~\ref{SDdet} where one of cuts is $\sum E_T^{\rm fw} < 6$~GeV.
 It is also evident that the pile-up effect is slightly more pronounced for SD background than for
 the signal.

 As demonstrated, however, on the zero pile-up case in Fig.~\ref{fig:stdet60} (top left), the effect of the  
 detector-level $\xi^{\rm calo} < 0.025$ cut is remarkable -- it brings the S/B ratio back above 1.0. Thanks to the
 much bigger resemblance of the SD background
 to the signal, it is the SD process which dominates the total contamination. The situation dramatically changes when
 combinatorial effects enter the game: here the ND background dominates
 thanks to roughly two orders of magnitude higher cross section than the SD background and to the fact that the
 size of the combinatorial factor $P_{\rm comb}$ stays the same for both ND and SD backgrounds. The mean value
 of the $\xi^{\rm calo}$ distribution rises with increasing $\langle \mu \rangle$, but more for signal than for
 both backgrounds. Convoluted with the increasing combinatorial factor $P_{\rm comb}$ this leads to a decreasing S/B
 ratio as the level of pile-up increases. 

 The final event yields for signal, SD and ND dijet backgrounds are shown in
 Table~\ref{tab:sig}. As explained above, they correspond to doubling those obtained from analyzing
 all generated event samples since the signal has been generated with the forward proton scattered
 in one hemisphere only and consequently the selection cuts have been tailored to one hemisphere
 as well.

For the specific cut scenarios in Eq.~(\ref{Eq:goldendet60}) and Eq.~(\ref{Eq:goldendet100}) considered for the two
mass intervals, we observe that the S/B ratio safely exceeds unity when pile-up is not considered, thanks to the
efficient $\xi^{\rm calo}$ cut. If we require $\xi^{\rm calo} < 0.025$, the S/B stays above unity for
$\langle \mu \rangle = 1$ and $M_{\rm inst} > 60$~GeV, all other luminosity scenarios give S/B below 1.0. Its value
further decreases with increasing $\langle \mu \rangle$ and reaches minima of 0.3 (0.2) at $\langle \mu \rangle = 5$
for $M_{\rm inst} > 60$ (100)~GeV. It is therefore a clear preference to collect data at rather
low amounts of pile-up, and we believe that a special run with $\langle \mu \rangle \sim$~1 and $\cal L \sim$~0.1
fb$^{-1}$ is realistic to consider.
 \begin{table*}[h]
   \begin{center}
     \begin{tabular}{|c|c|c|}       
  \hline
  ($\langle \mu \rangle, \cal L$[fb$^{-1}$]) & $M_{\rm inst} > 60$~GeV & $M_{\rm inst} > 100$~GeV \\ \hline
  (0, 0.1)   & 19.0/(0.4+3.5)  & 5.8/(0.2+3.5)   \\ \hline
  (1.0, 0.1) & 8.7/(6.5+0.2)  & 3.2/(4.7+0.2)   \\ \hline  
  (2.0, 1.0) & 52.2/(58.1+2.5) & 15.4/(55.3+2.2) \\ \hline 
  (5.0, 10.0)& 56.2/(205.6+13.3)& 23.8/(137.1+7.6)\\ \hline  
\end{tabular}
\vspace*{0.3cm}
\caption{Summary of event yields after applying cuts in Eq.~(\ref{Eq:goldendet60}) and 
  Eq.~(\ref{Eq:goldendet100}) for the single-tag search approach for $M_{\rm inst} > 60$~GeV
  and $M_{\rm inst} > 100$~GeV, respectively, and for four luminosity scenarios ($\langle \mu \rangle$, $\cal L$).
  For each scenario, a ratio of number of signal to background events, $N_{\rm S}/(N_{\rm ND}+N_{\rm SD})$, is shown.}
\label{tab:sig}
\end{center}
 \end{table*}
\subsection{Systematic studies or searching for an optimum working point}
In the effort to improve S/B ratios, we investigate variations of the selection cuts or tracking
efficiency. We observe very similar numbers for a loosened cut $\sum E_T^{\rm fwcalo} < 6$~GeV.
As anticipated, we also study the effect of a more efficient tracking. This leads to an increase
of the average efficiency from 80\% to 90\% and hence of the track multiplicity. This allows us to apply
a more strict cut on the track multiplicity for $M_{\rm inst} > 60$~GeV, namely $N_{\rm tr05} > 28$, but
the S/B ratios appear to be unchanged. Tightening the track multiplicity cut to $N_{\rm tr05} > 26$ and
$N_{\rm tr05} > 27$ for $M_{\rm inst} > 60$~GeV and the original tracking or to $N_{\rm tr05} > 30$ for
the optimal tracking does not change the S/B ratio significantly either. For $M_{\rm inst} > 100$~GeV
tightening the track multiplicity to $N_{\rm tr05} > 32$ or $N_{\rm tr05} > 35$ increases the S/B ratio
with respect to that achieved for the scenario in Eq.~(\ref{Eq:goldendet100}),
however, since very few signal events survive, 
statistical significances would be rather small thus do not make this option viable.

The numbers in Table~\ref{tab:sig} correspond to the whole $S_T$ spectrum shown in Fig.~\ref{fig:stdet60}.
Restricting ourselves to the region where the instanton signal is expected, namely $S_T > 0.5$, would in general
lead to an increase of S/B by roughly 30\%, only for the highest pile-up point the increase is above 100\%. The
improvement of significances is rather modest.

 \subsection{Double tag}
As discussed in \cite{Khoze:2021pwd}, the central instanton production processes have some promising
advantages. Thus, in such a case, the Pomeron-Pomeron colliding energy is relatively low, 
which strongly reduces the multiplicity of the background underlying events. 
Moreover, in this energy range, the central detector becomes almost hermetic (close to $4\pi$) for
the Pomeron-Pomeron secondaries, and only a small part of the finally produced hadrons will avoid
detection.\footnote{The idea to observe instantons in a 2-Pomeron collision was first put forward
in Ref~\cite{Shuryak:2003xz} in the context of Pomeron collisions at very low invariant mass, 
$2 < M_{\rm inst} < 5$ GeV.}. Note also that detecting two outgoing protons would allow one to place an upper
limit on the instanton mass.

For the double-tag approach, we use the signal sample generated using Pomeron-Pomeron collisions.
We studied a limited set of cut scenarios which are close to the golden one (Eq.(\ref{Eq:goldenscenario}))
and observe selection efficiencies that are in the ballpark with those obtained on the proton-Pomeron
signal sample. Given about 80 times smaller production cross-section compared to the proton-Pomeron
case, that would give fiducial cross-sections of the order of fractions of femtobarns. We have 
to consider higher values of $\langle\mu\rangle$ with correspondingly larger integrated luminosities,
for example ($\langle\mu\rangle$, $\cal L$ [fb$^{-1}$]) = (20, 60) and (50, 300). This would lead
to increasing the event yields for all, signal and backgrounds, by a factor of 60 or 300 with respect
to e.g. the (2,1) scenario considered in the ST approach. 

As explained in Section~\ref{sec:comb}, the combinatorial factor $P_{\rm comb}^2$ from the fake protons
seen in the acceptance of FPDs due to additional pile-up collisions is not large, due, mainly, to the narrow $\xi$
range chosen. It can then be further reduced by utilizing ToF detectors. The ToF detectors
with an assumed 20~ps resolution (which is a goal of AFP ToF in Run~3 and slightly better than
achieved by AFP ToF in Run~2~\cite{ToFPUBNote}) would be able to suppress the combinatorial background
by a factor of 9 and 8 for $\langle\mu\rangle$ of 20 and 50, respectively.
Taking into account the values of
$P_{\rm comb}^2$ evaluated above (0.84\% and 4.58\% for $\langle\mu\rangle$ of 20 and 50,
respectively), we can make a very rough extrapolation from the ST results discussed in the previous section
to much higher pile-up amounts considered for the DT analysis. If we for example compare the luminosity scenarios
(1,0.1) for ST and (20,60) for DT, the fiducial cross section of ND and SD backgrounds would be scaled by a 5 times
lower factor (0.84\%/9 vs. 0.48\%) for DT but the ratio of signal cross sections is 1/80. This leads
to a reduction of the S/B ratio obtained for the (1,0.1) scenario for ST by a factor of about 16. For the (50,300)
scenario in DT, the S/B ratio would even drop by two orders of magnitude. It should be noted that outlooks for
significances
may be more favourable since in both luminosity scenarios for DT we collect much more statistics than for ST but
it has to be added that the expected effect of such a high level of pile-up on both, signal and backgrounds, is
a considerable drop of selected events. A central instanton production either in the DT or ST approach deserves
a detailed analysis including the simulation of detector and pile-up effects, and is in our plans for a future
paper.

\subsection{Potential improvements}
Triggers were not discussed in this note and to our knowledge, there are no dedicated instanton triggers
being used by the LHC experiments, so instanton signals would have to be searched in data collected by
other triggers, e.g. very-low $E_T$ jets or minimum bias triggers used to study properties of charged
particles. It would, however, be extremely useful to propose and build triggers with conditions
tailored to collect instanton-enhanced data samples, e.g. those in
Eqs.~(\ref{Eq:goldenscenario},~\ref{Eq:goldendet60},~\ref{Eq:goldendet100}).
So far tracking information is not available at L1 and so are not vertices since vertex reconstruction is a
time-consuming procedure but triggering an instanton-like signal already at L1 would make the instanton
search in collected data more efficient. At HLT, a full-scan tracking (collecting information about track
multiplicities, $\eta$, $p_T$ or $\sum E_T$) can be obtained in special low pile-up runs. In standard runs
other approaches should be taken, for example to run the full-scan online tracking in coordination
with very low-$E_T$ jet or other-soft scale triggers, getting possibly also the full information about
primary vertices. Tracking in a limited region-of-interest (similar to that in Eq.~(\ref{Eq:goldendet60})
or Eq.~(\ref{Eq:goldendet100})) should be possible at HLT too. 

As we have discussed in Section~\ref{SDdet}, pile-up increases enormously $\sum E_T$ in the forward
region ($2.5 < \lvert\eta\rvert < 4.9$), according to Delphes with the ATLAS input card, from 9~GeV at zero
pile-up up to 37~GeV at $\langle\mu\rangle$ of 5 for signal. This is because the tracking information
is missing there so we do not have control over what fraction of energy comes from pile-up. One can
of course require just one primary vertex which would greatly suppress pile-up effects but would
drastically reduce available statistics which is not affordable. Alternatively, adding time information
of individual calorimeter cells, if possible already in Run~2 and Run~3, should help to identify those
coming from pile-up vertices distant from the primary one. This should be
much improved in Run~4 where both, trackers in ATLAS as well as CMS will be upgraded to cover a region
$\lvert\eta\rvert < 4.9$ and time information of tracks should be available as well~\cite{CMSMIP,ATLASHGTD}.
This time information about tracks in the central detector, whether alone or together with the ToF information
about the leading forward proton on one side from the interaction point will not only help in searches
for the proton-Pomeron signal but could also potentially allow us to use 
single-tagged events to separate the Pomeron-Pomeron-induced instanton signal from all backgrounds
(in this case the undetected proton will often have $\xi < 0.02$, below the current FPD acceptance).
Indeed this combined time information and the increased selection efficiency expected for single-tagged
events promises to significantly increase the S/B ratio compared to double-tagged approach for the
search for instantons produced in the Pomeron-Pomeron collisions.

Next, we can exploit the fact that the density of secondary particles ${\rm d}N_{\rm ch}/{\rm d}\eta$ reaches its
maximum at the rapidity $\eta=\eta_{\rm inst}$ equal to the instanton rapidity. Indeed, in terms of $\eta$
the spherical distribution caused by the instanton decay reads

$$ {\rm d}N_{\rm ch}/{\rm d}\eta\propto 1/\cosh^2(\eta-\eta_{\rm inst}). $$

In the present study we chose the interval $0<\eta<2$ which, on average, corresponds to this maximum,
however, the statistical significance of the result may be improved if the $N_{\rm ch}$ and $\sum E_T$ cuts
are imposed in the central detector within the $\eta_0<\eta<\eta_0+2$ interval, where the particle density is
maximal in {\em each} particular event. We are planning to implement this idea in future studies.

Finally we remind that the decay of the instanton produces one additional pair of each flavour of light
($m_f<1/\rho$) quark~\cite{Khoze:2019jta,Amoroso:2020zrz,Khoze:2021jkd}. So in the case of the signal, we expect to observe a
larger number of strange and charm particles than in background events. While this fact has not been
examined in this study, we believe it has potential to improve the S/B ratio.

\section{Conclusions}\label{sec:conclusions}
In Ref.~\cite{Khoze:2021jkd,Khoze:2021pwd} it was proposed to search for QCD instantons at the LHC in events with
Large
Rapidity Gaps. Here we investigate the search strategy for the case of heavy instanton by tagging leading
protons with dedicated AFP or CT-PPS forward proton detectors at $\sqrt s = 14$~TeV. Since the expected cross-sections
are quite small
we consider scenarios with relatively large integrated luminosities and, thus, we have to account for
effects of pile-up. In addition, we include fast simulation of central detector response. For instanton
signal, we use a dedicated MC event generator RAMBO, while the dominant backgrounds, Single-diffractive
and Non-diffractive dijets are estimated using PYTHIA 8.2, all including full underlying event simulation.
The small size (heavy) instanton produces a large number of mini-jets and large transverse energy in a
limited rapidity interval, therefore we are selecting events with a large multiplicity. To suppress the main
background caused by multiple parton interactions and pile-up events, we introduce an additional cut,
transverse energy in the forward ($2.5<\eta<4.9$) calorimeter should be less than 5--6~GeV. The combinatorial
background, caused by fake protons from pile-up interactions seen in the acceptance of FPD, is greatly
reduced if we limit the acceptance interval to $0.02 < \xi < 0.05$. We investigate two proton tagging strategies,
the single-tag and double-tag with appropriate initial collisions, namely proton-Pomeron and Pomeron-Pomeron,
respectively, both simulated by RAMBO. By concentrating on instanton masses larger than 60~GeV, we show that
by applying appropriate cuts and requiring one leading proton, we could expect the signal-to-background ratio
S/B $> 2.3$ at generator level. To keep the pile-up effects in both, the central and forward parts of the
main detector at a tolerable level, we choose to work at a relatively low pile-up rate, with
$\langle\mu\rangle < 5$. We show that the detector and pile-up effects are kept under control and consequently
    the S/B ratio kept well above unity if $\langle\mu\rangle \lesssim 1$ and integrated luminosity is around
    0.1 fb$^{-1}$.

The kinematics with two tagged leading protons have some advantages. In this case, the central detector
becomes almost hermetic for the Pomeron-Pomeron collision. Unfortunately, the expected cross-section
becomes about 80 times smaller compared to proton-Pomeron collisions. That is, we have to consider a
larger $\langle\mu\rangle =$~20--50. In such a case, even with a good forward proton timing resolution
the combinatorial background caused by the non-diffractive events, accompanied by leading protons
from two other soft events turns out to be too large to reach advantageous S/B ratios, although a detailed
analysis including detector and pile-up effects is necessary to make firm conclusions.

For any instanton searches in both, the proton-Pomeron and Pomeron-Pomeron collisions, the additional time
information about tracks in the central and forward rapidities seems to be beneficial.

\section*{Acknowledgments}
Computational resources were provided by the Computing Center~\cite{Adam:2020qor} and by CEICO (Central
European Institute of Cosmology) both at the Institute of Physics of the Czech Academy of Sciences.
The authors are also grateful to Valya Khoze for valuable discussions and encouragement.
DLM is supported by an STFC studentship.
MT is supported by the Ministry of education, youth and sport of the Czech Republic within the project LTT17018.

\bibliography{inst}

\end{document}